\documentclass{optica-article}

\journal{opticajournal} 

\articletype{Research Article}
\usepackage{hyperref}
\hypersetup{colorlinks,linkcolor={blue},citecolor={blue},urlcolor={red}}  

\usepackage{rotating}

\begin{document}

\title{High-efficiency, high-count-rate 2D superconducting nanowire single-photon detector array}

\author{Fiona Fleming,\authormark{1,*} Will McCutcheon,\authormark{1} Emma E. Wollman,\authormark{2} Andrew D. Beyer,\authormark{2} Vikas Anant,\authormark{3} Boris Korzh,\authormark{4} Jason P. Allmaras,  \authormark{2} Lautaro Narv\'{a}ez,\authormark{5} Saroch Leedumrongwatthanakun,\authormark{1,$\dagger$} Gerald S. Buller,\authormark{1} Mehul Malik,\authormark{1} Matthew D. Shaw\authormark{2}  }

\address{\authormark{1}Institute of Photonics and Quantum Sciences, School of Engineering and Physical Sciences, Heriot-Watt University, Edinburgh, EH14 4AS, UK\\
\authormark{2}Jet Propulsion Laboratory, California Institute of Technology, Pasadena, CA 91109, USA\\
\authormark{3}Photon Spot, Inc., 142 W. Olive Ave., Monrovia, CA 91016, USA\\
\authormark{4}Group of Applied Physics, University of Geneva, 1205 Geneva, Switzerland\\
\authormark{5}Division of Physics, Mathematics and Astronomy, California Institute of Technology, Pasadena, California 91125, USA\\
\authormark{$\dagger$}Current address: Prince of Songkla University, Thailand}

\email{\authormark{*}f.fleming@hw.ac.uk} 

\begin{abstract*} 
Superconducting nanowire single-photon detectors (SNSPDs) are the current leading technology for the detection of single-photons in the near-infrared (NIR) and short-wave infrared (SWIR) spectral regions, due to record performance in terms of detection efficiency, low dark count rate, minimal timing jitter, and high maximum count rates. The various geometry and design parameters of SNSPDs are often carefully tailored to specific applications, resulting in challenges in optimising each performance characteristic without adversely impacting others. In particular, when scaling to larger array formats, the key challenge is to manage the heat load generated by the many readout cables in the cryogenic cooling system. Here we demonstrate a practical, self-contained  64-pixel SNSPD array system which exhibits high performance of all operational parameters, for use in the strategically important SWIR spectral region. The detector is an 8x8 array of 27.5 $\times$ 27.8 µm pixels on a 30 µm pitch, which leads to an 80 -- 85\% fill factor. At a wavelength of 1550~nm, a uniform average per-pixel photon detection efficiency of 77.7\% was measured and the observed system detection efficiency (SDE) across the entire array was 65\%. A full performance characterisation is presented, including a dark count rate of 20 cps per pixel, full-width-half-maximum (FWHM) jitter of 100 ps per pixel, a 3-dB maximum count rate of 645 Mcps and no evidence of crosstalk at the 0.1\% level. This camera system therefore facilitates a variety of picosecond time-resolved measurement-based applications that include biomedical imaging, quantum communications, and long-range single-photon light detection and ranging (LiDAR) and 3D imaging.
\end{abstract*}

%%%%%%%%%%%%%%%%%%%%%%%%%%  body  %%%%%%%%%%%%%%%%%%%%%%%%%%
\section{Introduction}
In recent decades, SNSPDs have emerged as remarkably high-performance photon counters, demonstrating record levels of efficiency, jitter performance, dark count rate, dynamic range, and operational wavelength range \cite{Zadeh}. As a result, these detectors have found particularly high prevalence in applications which exploit their picosecond time resolution and high detection efficiency such as LIDAR \cite{Taylor:20,McCarthy:13}, 3D imaging\cite{Hadfield:23}, quantum experiments with single photons\cite{Krenn_2016} and long range free-space communications\cite{Wollman2024}. The first demonstration of a superconducting wire strip utilised as a photon sensor was reported by  Gol’tsman et al.~in 2001 \cite{Goltsman2001}. Since then, the field of SNSPDs has experienced rapid growth due to improvements in their performance characteristics and a vast and growing number of potential application areas. This has led to the development of SNSPD-based systems with detection efficiencies greater than 95\% \cite{Reddy:20, 10.1063/5.0039772, Hu2020}, jitter as low as < 3~ps\cite{Korzh2020}, high count rates reaching 1.5 Gcps \cite{Craiciu:23,Resta2023,Hao2024,Zhang2024} and dark count rates down in the micro-Hertz range\cite{PhysRevLett.128.231802}. 

The development of a high-efficiency, low-jitter SNSPD array is particularly exciting for the rapidly advancing fields of high-dimensional quantum photonics and quantum imaging \cite{MalikBoyd_2014, Padgett_2019, Defienne_2024, Krenn_2017, Cozzolino_2019}. In the former, single-pixel detectors are commonly used in combination with wavefront-shaping devices such as spatial light modulators (SLMs) to perform generalized projective measurements of SWIR single-photon spatial modes \cite{Valencia2021,Srivastav2022,kopf2024}. However, such an arrangement can only be used to measure one spatial mode at a time, which severely limits the practical implementation of high-dimensional quantum information protocols that include noise and loss-robust entanglement distribution \cite{Srivastav2022} and high-capacity quantum key distribution \cite{Mirhosseini_2015}. A high-efficiency SWIR SNSPD imaging array with even tens of pixels would serve to significantly advance this field. Combined with state-of-the-art techniques for mode transformation such as complex media-based circuits \cite{Goel2024-kq} and multi-plane light convertors (MPLCs) \cite{Brandt_2020,Lib_2022, goel2022simultaneously}, an SNSPD array would enable generalized multi-outcome measurements of photonic spatial modes that would find application in state-of-the-art quantum communication and imaging protocols. Such a device would also be particularly useful in fundamental studies of high-dimensional entanglement \cite{Erhard_2020}, where recent demonstrations include a single-outcome measurement of 55-dimensional “pixel” entanglement \cite{HerreraValencia2020} and the use of a time-stamping camera for certifying 14-dimensional entanglement \cite{Courme:23}. There are significant advantages of operating a SWIR single-photon detector array for various other free-space classical and quantum communication applications, notably in LiDAR measurements. The laser eye-safety threshold at these wavelengths is greater than at shorter wavelengths resulting in improved signal-to-noise and/or longer operation ranges\cite{Wallace:2020}. Transmission through the atmosphere is increased due to less scattering from atmospheric obscurants \cite{Tobin2021-ou,Arnulf:57,Tobin:19}and the solar background is reduced in the SWIR\cite{BIRD1983563}, again leading to an improved signal-to-noise ratio in quantum communication and photon-starved LiDAR measurements. Gas sensing also benefits from detector arrays able to target different absorption peaks across a broader range of wavelengths\cite{Mackenzie2023}.

 In this work, we present the characterisation and operation of a practical 64-pixel SNSPD camera system for picosecond time-resolved imaging. The detector cryostat and electronics are contained within a mobile electronics rack which provides ease of user access for operation (see supplementary material). To our knowledge, the array presented is the most optimally performing time-correlated single-photon counting (TCSPC) imaging SNSPD array to date in the SWIR range, with a particularly high SDE and low timing jitter for this pixel count. We choose a pixel size of $\sim30$~\textmu m to maximize the fill factor and consequently the array system detection efficiency (SDE) (65\%), while simultaneously achieving a sufficiently low timing jitter ($\approx$100 ps full-width at half-maximum (FWHM)) and dark count rate (20 cps/pixel) suitable for quantum photonics experiments. This system represents a significant improvement over early proof-of-principle imaging arrays \cite{10.1063/1.4921318} and demonstrates a high-performance, self-contained, and versatile detector system which is ideally suited for a range of advanced applications. 
While our system is specifically designed for experiments on high-dimensional quantum photonics \cite{Xu:16, Srivastav2022,Goel2024-kq, lib2023}, it would also be particularly beneficial in 3D imaging and LiDAR applications requiring high spatial and temporal resolution\cite{Tachella2019-hm} with the potential for imaging of cluttered scenes\cite{10.1117/1.NPh.8.3.035006,Maccarone:19}.

\section{Experimental Setup}

When scaling to large detector array formats, one of the main challenges is effective management of the heat load from the high-speed coaxial cables required for biasing and signal readout. The most straightforward method of addressing an array of SNSPDs is direct readout, where each pixel has its own individual readout channel. This technique becomes impractical as the number of channels increases, because each RF readout cable imposes a heat load on the cryostat. Therefore, many cryogenic multiplexing techniques have been proposed in attempts to reduce  the thermal strain on the system\cite{Doerner2017, Miki:14, Yabuno:20,Gaggero:19, Zhu2018-qn, 10.1063/1.4921318, Wollman:19,Allmaras2020,ZHANG2023100056,Zhao2017-aq}. To date, the largest SNSPD array demonstrated has 400,000 pixels, which are addressed using a combination of thermal row-column and time-domain multiplexing techniques\cite{Oripov2023-vf}. The various multiplexing architectures usually involve a tradeoff between increasing pixel number and reducing performance in other metrics. In particular, these architectures face challenges with handling high count rates or multi-photon events. Furthermore, pixel fill-factor and pixel-to-pixel crosstalk can prove challenging for SNSPD imaging arrays, whether multiplexed or addressed via direct-readout.
 Direct readout remains the most straightforward and well-established technique to achieving high system performance across many metrics simultaneously.
 
\begin{figure}
\centering
\includegraphics[width=1\linewidth]{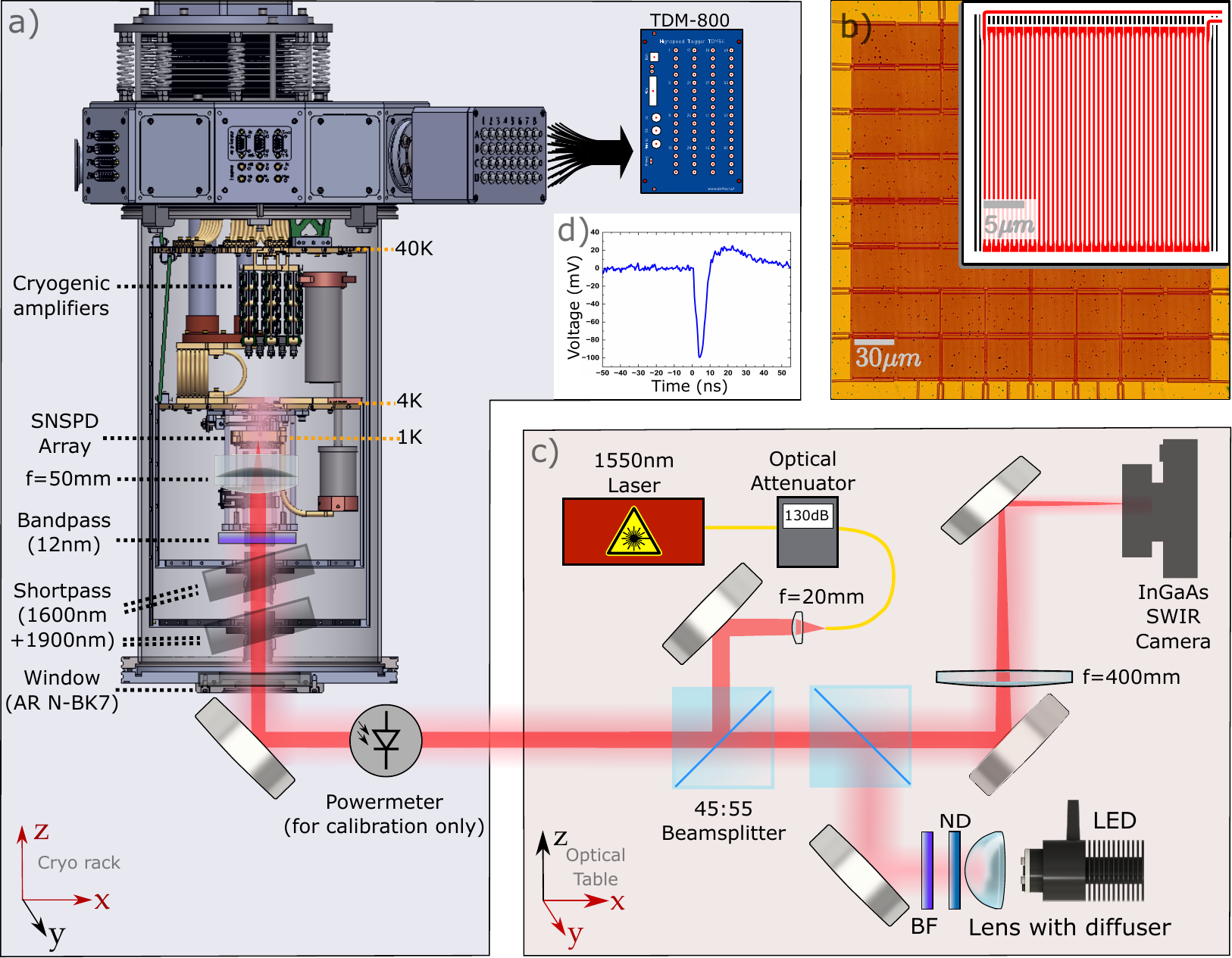}
\caption{\label{fig:set-up}Schematic diagram of the optical alignment and imaging system for the 64-pixel SNSPD array: a) Cryogenic rack internal components, optics, and timing electronics (TDM-800). b) Optical micrograph of the SNSPD array and svg of single pixel nanowire (inset). c) Schematic diagram of the array illumination and imaging system. d) An example output voltage pulse from an individual SNSPD pixel within the array. This pulse was measured with a bias current of 21 $\mu$A applied to the nanowire and is representative of pulses from each of the individual 64 channels.}
\end{figure}

 A schematic diagram detailing the experimental setup employed in this work is shown in Fig.~\ref{fig:set-up}. The cryogenic rack arrangement and components can be seen in Fig.~\ref{fig:set-up}a. The SNSPD array is designed in a direct readout format and is fabricated from NbTiN with a gold back mirror and a bi-layer anti-reflection coating. The AR coating and back reflector form a low-quality-factor optical cavity centered at a wavelength of 1550~nm in order to enhance optical absorption into the nanowire layer. The nanowires are 250~nm wide with a 555~nm pitch. Outside of the pixel area, the nanowires widen to 380~nm until they are routed outside the array area. The nanowires in each pixel are meandered to cover an active area 27.8~\textmu m wide and 27.5~\textmu m tall. However, the bends of the meander are widened to prevent current crowding, and therefore may have a lower efficiency than the rest of the pixel. Excluding the bends, the pixel active area is 26.1~\textmu m tall. The pixel pitch is 30~\textmu m, resulting in a fill factor between 80 -- 85\% (Fig.~\ref{fig:set-up}b). The array is packaged in a pin grid array (PGA) chip carrier for easy screening on a testbed cryostat using a zero-insertion force PGA socket, then installed in a low insertion force PGA socket on an interface PCB in the operational cryostat. Each pixel in the array is wired to a pair of bond pads on the chip to support differential readout\cite{8627944}, which can reduce geometric jitter. However, to simplify the readout system, one end of each wire is grounded at the interface PCB, so the array can be read out using 64 standard coaxial cables. The SNSPD array is cooled to a temperature of 920~mK using a pulse-tube cryocooler and a helium sorption refrigerator. Eight 8-channel amplifier boards at 40~K amplify the detector pulses and contain bias tees for low-noise biasing of the pixels. The amplified pulses are read out using a 64-channel time-to-digital converter with front-end comparator (Dotfast TDM800-64). The TDC registers the arrival time of the pulses from the SNSPDs with $15.625$ ps resolution, $20$ ps RMS jitter, and up to a $900$Mcps tag rate. Room-temperature DAC modules are used to apply bias currents to the pixels. The location of the array and electronics inside the cryostat is shown in Fig.~\ref{fig:set-up}a.

The arrangement on the optical table for coupling laser light onto the detector is shown in Fig.~\ref{fig:set-up}c. A 20 mm focal length lens outside the cryostat collimates the laser light from a single-mode optical fibre. The collimated beam is first directed through a 50/50 beamsplitter in the horizontal plane (x,y axes) and then along the vertical plane (z,x axes) using a dielectric folding mirror. The x-y position of the detector can also be adjusted relative to the cryogenic lens mount. A broadband LED illumination source with a central wavelength of $\approx$1550 nm and an aspheric condenser lens are used to flood-illuminate the array. The LED output is directed onto the array by another beamsplitter and folding mirror at the cryostat entrance and then imaged back onto the SWIR camera as shown. The outer cryostat housing is fitted with an anti-reflective-coated N-BK7 window. Both the inner 4K and outer 40K radiation shields have 1600~nm and 1900~nm short-pass wavelength filters placed in the incident beam path which block room-temperature blackbody radiation from reaching the detector and causing false counts \cite{Mueller:21}.  In addition, a 12 nm bandpass filter centred at a wavelength of 1550~nm is contained inside the 4K shield compartment. Each filter contributes a few percent of loss at 1550~nm, so the array SDE can be improved by removing filters at the expense of additional dark counts. The focusing lens inside the cryostat is a 50 mm focal length, 25.4~mm diameter, anti-reflective-coated, achromatic doublet. The free-space filtering arrangement employed here can achieve high coupling efficiency whilst maintaining low dark count rates \cite{Mueller:21}. 

For the efficiency, count rate, and crosstalk measurements, a continuous wave (CW) laser (Santec wavelength selectable WSL-100) centered at a wavelength of 1550 nm was used. For the jitter measurements, a mode-locked ultrafast fibre laser (Calmar laser Mendocino FPL-01CFF) at 1550 nm wavelength with pulse duration of < 0.5 ps was used. An inline fibre attenuation system was used to control the input photon flux incident on the array to maintain a level on the order of $\approx$ \(10^6\) photons/s---this ensured the time tagger was operating below its maximum data rate. A  manual fibre polarisation controller was used prior to the fibre attenuator, to facilitate testing of the efficiency polarisation dependence.

\section{Results and Discussion}

An SNSPD is maintained in its superconducting state as long as the temperature is maintained below its inherent critical temperature ($T_{c})$, and any applied bias current is less than the switching current ($I_{sw}$) value. When a photon is incident on an SNSPD, the local superconductive state of the absorption site is disrupted, creating a resistive hotspot. The fast, measurable current pulse generated at this hotspot is shunted to the readout circuity, containing a resistor and the cryo-amplifiers, allowing the output voltage pulse to be measured. An example output pulse from one individual SNSPD pixel within the array is shown in Fig.\ref{fig:set-up}d. This pulse was recorded with a bias current of 21 $\mu$A applied to the nanowire. The kink in the falling edge of the pulse is likely due to reflections off of the end of the pixel that is grounded at the interface PCB. The slight positive overshoot following the initial pulse is due to high-passing of the SNSPD pulse by the cryogenic amplifiers. The actual electrical decay time of the SNSPD is longer than the pulse width would indicate.

\begin{figure}
\centering
\includegraphics[scale = 0.63]{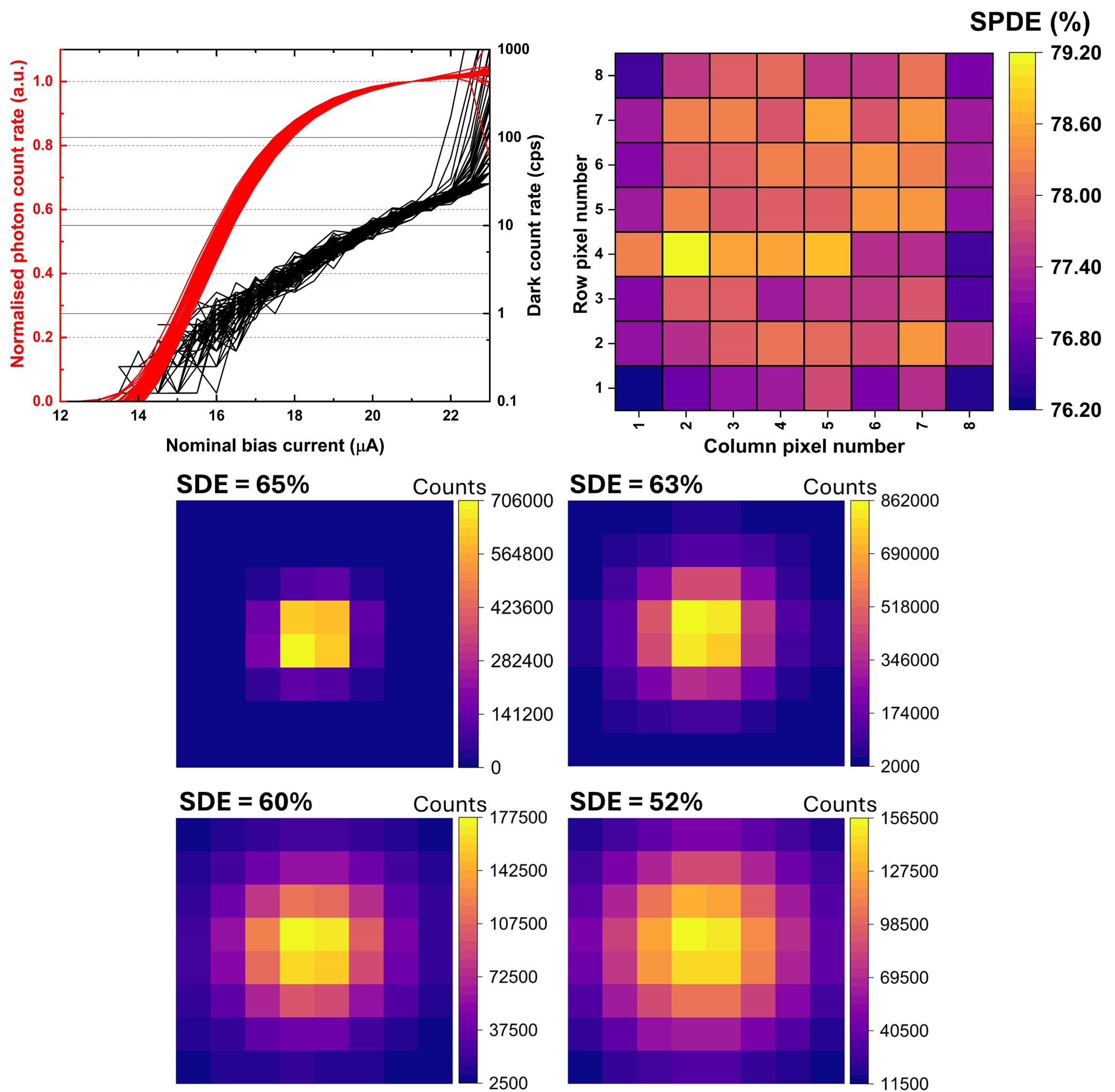}
\caption{\label{fig:PCR} a) Graph of normalised photon count rate (red) and dark count rate (black) as a function of nominal bias current applied simultaneously to all 64 SNSPD detectors.b) Diagram of pixel layout and corresponding single-photon detection efficiency. All measured SPDEs have an error value of $\approx$ $\pm$5\% associated with the power meter uncertainty and the error in aligning the laser spot onto each pixel consistently. The pixels along the edges of the array are measured to have lower efficiency, because some of the optical power falls outside the array active area when the spot is centered on these pixels. c) Colour heatmap diagrams of the laser beam expanded over the 64 pixel SNSPD array.}
\end{figure}

The normalised photon count rate (PCR) and dark count rate (DCR) as a function of bias current applied simultaneously to all 64 channels is shown in Fig. ~\ref{fig:PCR}a) in red and black respectively. The PCR was measured with the array flood-illuminated by the CW 1550 nm wavelength laser, and the DCR was measured with the room-temperature cryostat window capped. All pixels show fairly uniform dark count rates with an exponential increase in dark counts occurring above $\approx$ 22 $\mu$A. For an applied bias current just below this exponential increase at 21 $\mu$A, the average DCR per channel is $\approx$ 20 cps resulting in an aggregate DCR of 1.28 kcps for the entire array. From the measurements in Fig.~\ref{fig:PCR}b, 21 $\mu$A was chosen as the nominal bias current to use for all other measurements.

Fig.~\ref{fig:PCR}b) shows the individual estimates of SPDE for each of the 64 pixels in the array. The SPDE was measured by using the folding mirror at the cryostat entrance to translate the focused laser spot across the array to maximize the count rate at each pixel. The SPDE for each pixel is calculated as ($R_{counts} - R_{dcr})/R_{input}$ where $R_{counts}$ is the total count rate recorded by the time tagger with the laser incident to the system, $R_{dcr}$ is the dark count rate (DCR), defined as the count rate recorded by the time tagger with no laser on and $R_{input}$ is the input photon flux. The main goal of the individual pixel measurements was to investigate the uniformity of the SPDE values across the whole array. $R_{counts}$ is taken as the total array count rate rather than the count rate measured for each individual pixel, this is because we know the pixel and TDC are not yet saturated at this applied bias. We can tell from the TDC output and the lens used that the laser spot overfills each pixel the individual pixel size of 27.8 $\times$ 27.5 $\mu$m. The   beam diameter of the approximated Gaussian focused spot is estimated to be $\approx$ 27 $\mu$m. If the spot was focused to be contained in a single pixel, we assume that the photons which are now absorbed in the nearest neighbouring pixels would be absorbed instead in the single pixel under measurement. Furthermore, the wire connections that traverse space in between the pixels are also part of the nanowire, and so any photons which are incident on them can also contribute to the total photon count - $R_{counts}$, but the wider wire width of the connection lines means that they have lower internal detection efficiency.  The input photon flux was measured by focusing the incident laser beam onto a free-space power meter placed prior to the cryostat folding mirror as shown in Fig.~\ref{fig:set-up}a). The mean average of the SPDE per pixel across the full array is 77.7\% with a standard deviation of $\sigma=\pm$0.6\%. However, the SPDE values measured have a significant estimated error value of $\approx$ $\pm$5\% due to both the power meter uncertainty and the error in aligning the laser spot onto each pixel consistently. The results in Fig.~\ref{fig:PCR}b) show a high degree of uniformity across the pixels with all SPDE values lying within $\approx$ 3\%. It is clear that the pixels near the edge and particularly the corners of the array have a reduced efficiency compared to the pixels nearer the centre of the array, this is due to the measurement method of $R_{counts}$ detailed above. SNSPDs are sensitive to the polarisation of the incident photons because of the directionality of the wires within a pixel. Light which is polarised parallel to the nanowires typically has higher absorption than that which is perpendicularly polarised\cite{Anant:08}. The measurements displayed in Fig.~\ref{fig:PCR}b) were carried out with the laser beam in the parallel-polarised configuration for maximum absorption. When perpendicular-polarisation was used, the mean average SPDE per pixel across the array is 73.7$\pm$0.5\%%. 

The maximum array SDE is defined as the fraction of photons in the input photon flux which result in a count collected by the time tagger. To measure this value, the laser beam was not accurately focused and positioned onto one pixel as in the Fig.~\ref{fig:PCR}b) measurements; it was instead positioned near the middle of the array and expanded by defocusing the collimation lens. The resulting images of the expanded spot are shown in Fig.~\ref{fig:PCR}c). The fill factor for this array is $\approx$ 80 -- 85\%, so the measured efficiency depends on the spot size -- very small spots centered in between pixels will have a lower efficiency due to the gaps between pixels/connecting wires, while very large spots will have a lower efficiency due to photons falling outside the array active area. The measured array SDE values corresponding to each spot size are also displayed, with the maximum value of 65\% being recorded for the spot size shown in the top left diagram of Fig.~\ref{fig:PCR}c). The heatmaps further demonstrate the uniformity of the efficiency between pixels in the array. This is a notably high value of efficiency across the full system. Despite the high SPDE values that single pixel SNSPDs are capable of achieving, often the architecture of a multipixel array combined with the other losses in the system, results in much reduced fill factor and efficiency values\cite{Miki:14,10.1063/1.4921318}. However, in this case the array and other sources of loss in the system, such as the filters and windows in the beam path (see Fig. 2) have been optimized to minimise optical losses, which in combination with the high fill factor helps to maintain a highly-efficient system. 

\begin{figure}
\centering
\includegraphics[scale = 0.39]{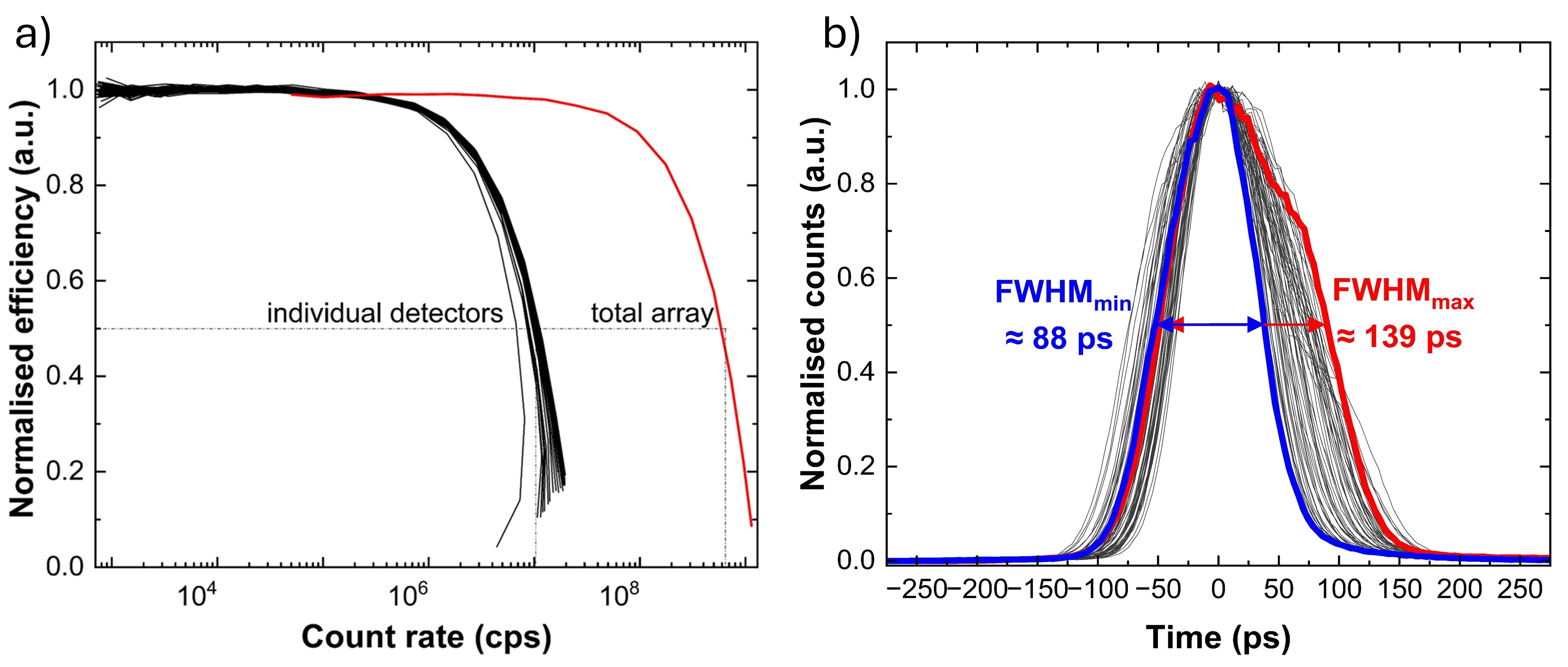}
\caption{\label{fig:mcr} a) Normalised efficiency of the 64 pixel SNSPD array versus measured count rate. The intersecting dashed lines indicate the MCR at the 3-dB compression point for the total array and also in the centre of the range of individual pixels. b) Temporal jitter histograms of all 64 pixel in the SNSPD array. The highlighted red and blue lines represent thefrom 2 individual pixels which exhibited the minimum (blue) and maximum (red) FWHM jitter values.}
\end{figure}

Maximum count rate (MCR) dictates the speed at which a detector can operate. High count rate operation is particularly beneficial for achieving fast acquisition times in single-photon LiDAR\cite{Hadfield:23}, and is also vital for other applications such as high data-rate quantum communications and QKD\cite{Dynes2016,Takesue2007}. To measure the MCR of the array, the CW laser spot was expanded to flood-illuminate the active area, and the count rate was measured as a function of laser power. The normalised efficiency of the entire array and for each individual pixel as a function of the measured count rate is shown in Fig.~\ref{fig:mcr}a). The total MCR of the array, taken at the 3-dB compression point as indicated by the dotted line in Fig.~\ref{fig:mcr}a), was measured to be $\approx$ 645 Mcps. This aligns well with the individual pixels' maximum count rates, which are mostly in the range from 8.2 - 11 Mcps. The MCR is currently limited by the electrical recovery time of the relatively large pixels - the same array design with smaller pixels would attain higher count rates, but at the expense of lower fill-factor. The scaling of maximum count rate with pixel number highlights one benefit of direct readout, as other mutliplexing schemes make a tradeoff between pixel number and maximum count rate. For example, the 400,000 pixel array reported in \cite{Oripov2023-vf} is limited to a count rate of 10 kcps, measured across a section of 50 detector sections in the array.

The temporal jitter of a photon-counting detector is defined by the uncertainty in a photon's arrival time as reported by the detector and time-tagging electronics and so is a measurement of the timing accuracy of the photon-detector system. SNSPD timing jitter is typically characterised as the full-width half maximum of the timing histogram obtained by illuminating the detector with a narrow bandwidth pulsed laser source. Lower jitter can result in improved signal-to-noise in LiDAR measurements, and reduced quantum bit error rate in quantum communications systems. The jitter measurement was done by forming a large spot across the entire array with a mode-locked ultrafast fibre laser. This $\lambda$ = 1550 nm laser had a  pulse width < 0.5 ps, and all detectors were biased simultaneously during the measurement. The TDC's comparator threshold for each pixel was chosen to minimize the jitter. One channel of the TDC was used to record the laser's electrical sync pulse, so two time-tag acquisitions were taken to capture the jitter of all 64 channels.  The jitter histograms were constructed from the difference in time between SNSPD events and the sync signal. The TDC's jitter introduces uncertainty in the sync signal timing, therefore, a weighted average over the preceding 8 sync periods was used to reduce but not eliminate the jitter introduced by the measurement. The TDC jitter ($\sim$50~ps FWHM) is also included in the timing of the SNSPD events. The graph highlights the timing histograms recorded from the two pixels which exhibited the minimum (blue line) and maximum (red line) FWHM values of the experimentally measured histograms. As can be seen in Figure~\ref{fig:mcr}b), most of the pixels demonstrated jitter values of $\approx$ 100 ps. The jitter is again limited by the large pixel size -- the jitter has a large geometric jitter contribution \cite{Korzh2020}, which is only partially cancelled out by choosing an optimal trigger threshold. In a system capable of handling twice the heat load, the double-ended design of the array could be used to cancel out the geometric jitter \cite{PhysRevApplied.19.044093}. The array jitter, although higher than for other SNSPD detectors at this wavelength, is still acceptable for many applications such as experiments on high-dimensional entanglement \cite{Erhard_2020}, quantum imaging \cite{Defienne_2024}, high-speed imaging\cite{Hadfield:23}, free-space optical communications\cite{Shaw2017}, and optical tomography\cite{Namekata2023-ky}.

\begin{figure}
\centering
\includegraphics[width=0.8\linewidth]{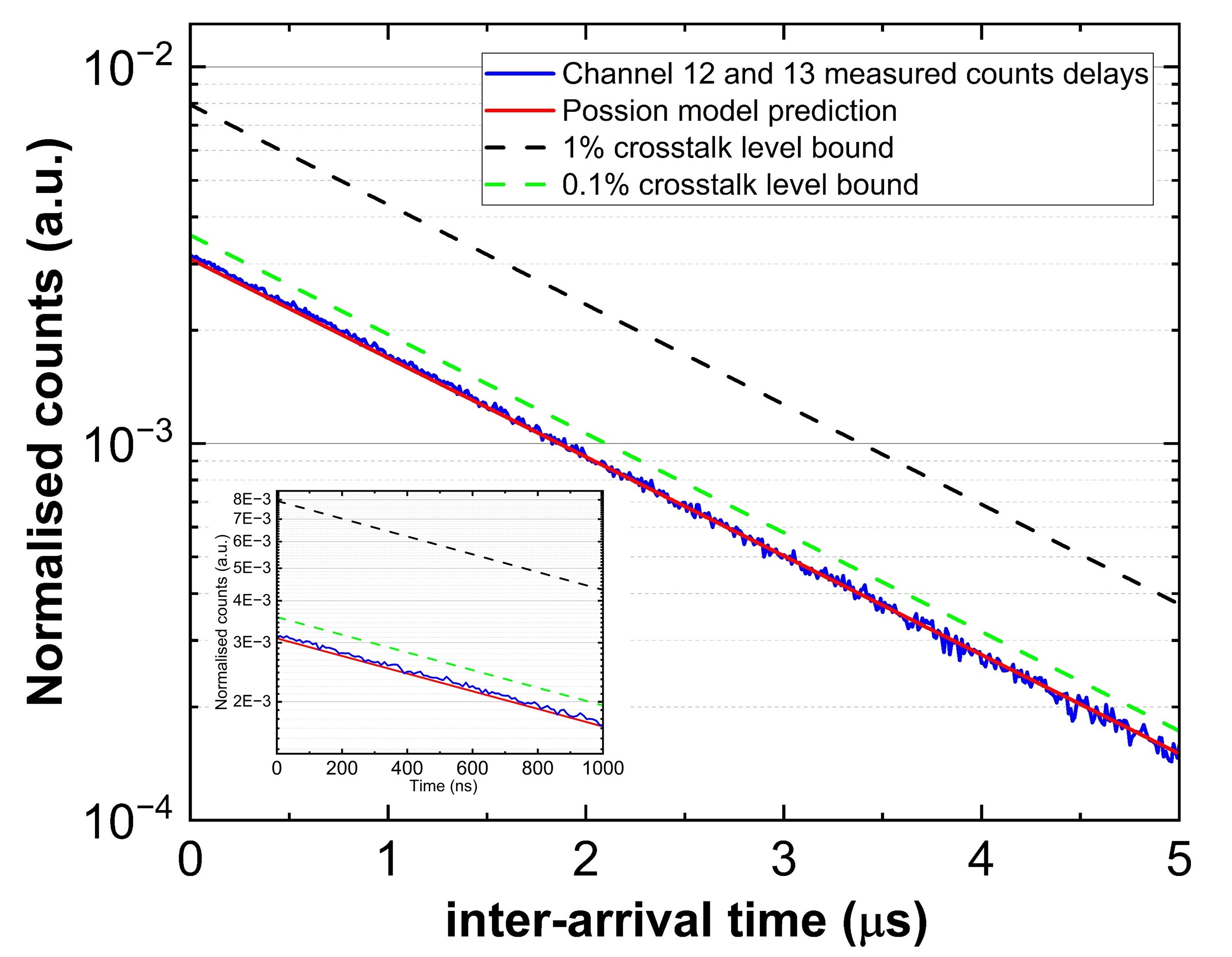}
\caption{\label{fig:crosstalk}Probability of detecting an event on Channel 13 following an event on Channel 12 as a function of inter-arrival time. Blue solid line: measured values. Red solid line: un-correlated Poisson model prediction. Black and green dashed lines: upper limit of correlations with 1\% and 0.1\% crosstalk respectively. The inset graph shows a zoomed in view on the timescale of 0 -- 1 $\mu$s. }
\end{figure}

Crosstalk in a photon counting array is an undesirable effect in which a detection event in one channel triggers a measurable event in another channel. The source of this crosstalk can be electrical or thermal. In the former, electrical coupling between channels can induce detectable pulses on channels which neighbour the channel in which the photon was absorbed. In an SNSPD array, thermal crosstalk occurs when heat generated at the resistive hotspot, formed at the photon absorption site, transfers to neighbouring channels and so disrupts the local superconductivity and triggers an event in those channels. Both crosstalk processes generally have a characteristic timescale, and therefore can be identified and measured by looking for correlations between events in different channels. If a single SNSPD pixel is illuminated with CW laser light, the output time-averaged count rate from that channel follows a Poisson distribution, and so counts from two independent channels should also follow Poisson statistics. The process used to analyse the crosstalk here is to observe the distribution of the inter-arrival time between events from pairs of channels, and to look for deviations from the exponential decay that would result from independent distributions. A crosstalk level of 1\% is defined as when 1\% of the total counts on one channel generate counts on a different channel. To test for this, the array was flood illuminated with the CW laser at 1550 nm wavelength whilst biased at 21 $\mu$A, and the time tags from all adjacent pairs of channels were analysed. With a pair of adjacent channels selected out, event pairs in which a count on the second channel followed a count by the first channel were identified. Then, the time delay between counts recorded on each channel was plotted, as shown in Fig. 6, where the normalised number of events where a count on Channel 12 preceded a count on Channel 13 are plotted vs. the time between the events, where Channels 12 and 13 correspond to 2 adjacent pixels within the array. The blue solid line is the measured data, the red solid line is the Poisson model prediction for zero crosstalk and the blue and green dashed lines are the upper limit of correlations corresponding to the 1\% and 0.1\% crosstalk levels respectively, the latter is included for comparison. From the typical thermal and electrical time constants involved in SNSPDs systems, a reasonable upper limit of 10 ns was chosen as the bin width for correlated events. Although this graph only shows the correlations for two particular nearest-neighbour channels, it is representative of the entire array. In Fig.~\ref{fig:crosstalk}, the observed interarrival time histogram lies along the line predicted by the Poisson distribution. Any counts due to crosstalk would appear as excess counts above the line. As seen in Fig. 5, crosstalk is statistically constrained below  the 30 dB (0.1\%) level.The inset graph shows a zoomed in view on the timescale of 1--1.8 $\mu$s and is included to show that there are no correlations on shorter timescales that may be hidden in the full plot. Based on the lack of observed correlations, it is therefore concluded that there is negligible evidence of crosstalk in the SNSPD array. 

One of the most widely used single-photon detectors in the SWIR spectral range are InGaAs/InP SPADs, which operate at Peltier-accessible temperatures $\approx$ 220 -- 255K. This results in considerably lower size, weight, power consumption and associated costs than cryogenic systems. They exhibit SPDE values in the SWIR of $\approx$ 15 -- 50\% \cite{Zhang2015-aq,Comandar2015-ki,Restelli2013-lj}. However, they tend to suffer from relatively high dark count rates and the deleterious effects of afterpulsing,  which typically limits the count rates attainable to 500 MHz in single-pixel SPADs\cite{ERFANIAN2024100617,mi14020437}. These drawbacks can make InGaAs/InP SPADs a sub-optimal choice for a SWIR single-photon detector array for some applications. Despite the unique challenges presented in designing a high-performance SNSPD array, the results presented here demonstrate that SNSPDs surpass InGaAs/InP SPADs in all of the above performance metrics apart from operating temperature and so provide a particularly high-performance alternative. Recently, the company Single Quantum has developed 36-pixel direct-readout SNSPD imaging arrays (with 24 connected pixels) optimised for operation at 1064 nm\cite{Tamimi:24} or 1550 nm wavelength\cite{10.1117/12.2692634}. These arrays have a pixel pitch of $\sim10$~\textmu m, an SDE of $\sim55\%$, an overall system jitter of < 50 ps, and a dark count rate of 150 cps per pixel. The commercialization of SNSPD imaging arrays shows that there is a demand for high-performing SNSPD arrays, with applications ranging from deep-space optical communications to fundamental studies of entanglement.

\section{Conclusion}
In summary, we present the fabrication and characterisation of a rack-mounted high-efficiency superconducting nanowire detector array system. The 64-pixel (8x8) array uses a direct readout scheme in which each pixel has its own individual readout line. The array showed very high uniformity across all 64 pixels in terms of pixel detection efficiency, with all 64 pixels achieving a maximum SPDE in the range 76.2 - 79.2\%, and a low dark count rate of 1.28 kcps across the entire array, which corresponds to $\approx$ 20 cps per channel. Furthermore, the maximum SDE of the overall system was measured as 65\%, a record high compared to other similar SNSPD arrays operating at 1550 nm. The array also exhibited a timing jitter of 100 ps per channel and a maximum count rate of 645 Mcps. There was negligible evidence of crosstalk between pixels. The high uniformity and performance of all parameters tested means that this system provides a laboratory-based SWIR single-photon sensitive detection system. Such a versatile system operating in this spectral range is highly suitable for experiments on high-dimensional quantum photonics, such as studies of spatial-mode entanglement \cite{Valencia2021, HerreraValencia2020}, multi-level quantum communication \cite{Mirhosseini_2015,Cozzolino_2019}, and quantum-enhanced imaging \cite{Defienne_2024}. Furthermore, it can be easily adapted for a range of free-space applications such as high resolution single-photon LiDAR imaging, especially in extreme environments such as long range\cite{Hadfield:23,Taylor:20,McCarthy:13}, through obscurants\cite{Tobin2021-ou}, and for imaging complex scenes\cite{Tachella2019-hm,Maccarone:19}. \\

\noindent\textbf{Acknowledgements.} We acknowledge funding from the European Research Council (ERC) Starting Grant PIQUaNT (950402), the Royal Academy of Engineering Chair in Emerging Technologies programme (CiET-2223-112), and the UK Engineering and Physical Sciences Research Council Centre-to-Centre grant (EP/W003252/1). Part of this work was performed at the Jet Propulsion Laboratory, California Institute of Technology, under a contract with the National Aeronautics and Space Administration (80NM0018D0004). Part of this work was performed at Caltech under the Alliance for Quantum Technologies INQNET framework.  

\bibliography{sample}

%% Input Supplemental Document
\newpage

\title{A high-efficiency, high-count-rate superconducting nanowire single-photon detector imaging array: supplemental document}

Here we show detailed photographs of the SNSPD array and cryostat mounted in a portable rack, alongside control, monitoring, and timing electronics, as well as close-ups of the array imaging system and chip carrier. 

\section{Photographs of the Cryostat system}

\begin{figure}[h!]
\centering
\includegraphics[scale = 0.7]{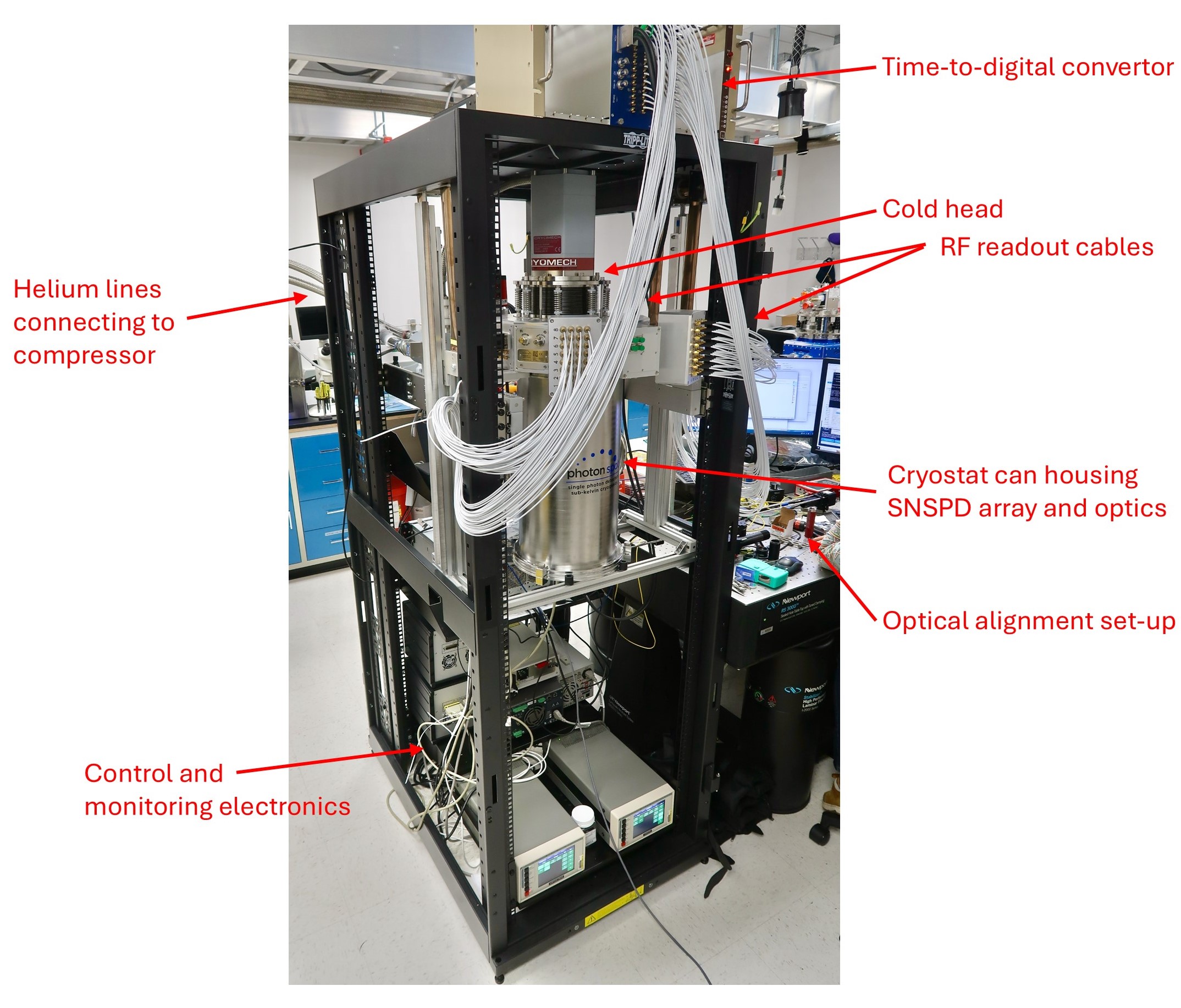}
\caption{\label{fig:rack}Photograph of rack containing the SNSPD camera cryostat, cryogenic control instruments, SNSPD electronics modules, and time-tagging electronics.}
\end{figure}

\begin{figure}[t!]
\centering
\includegraphics[scale = 0.3]{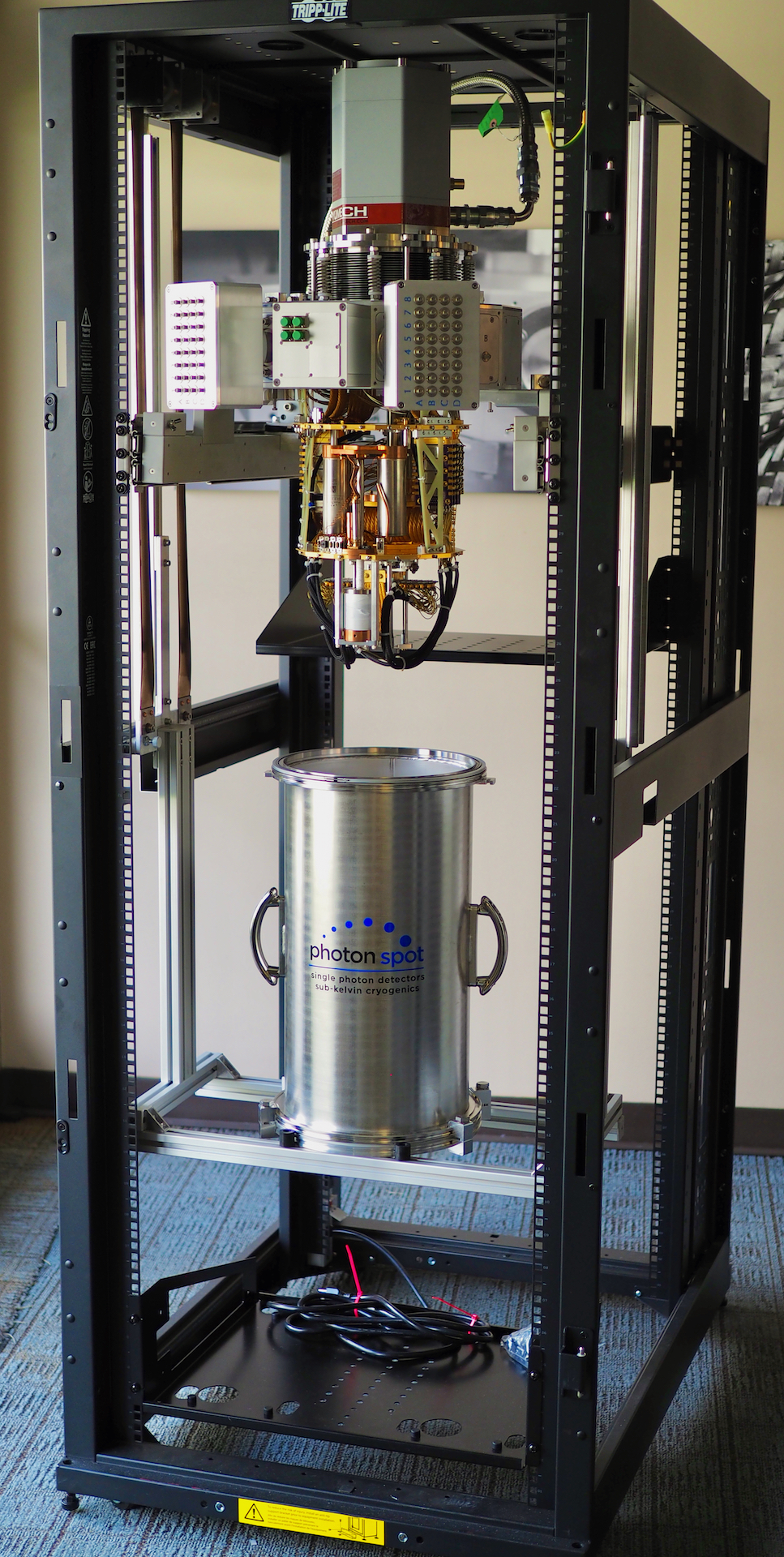}
\caption{\label{fig:rack1}Lowered cryostat housing and rack-mounted lowering mechanism.}
\end{figure}

\begin{figure}[t!]
\centering
\includegraphics[scale = 0.27]{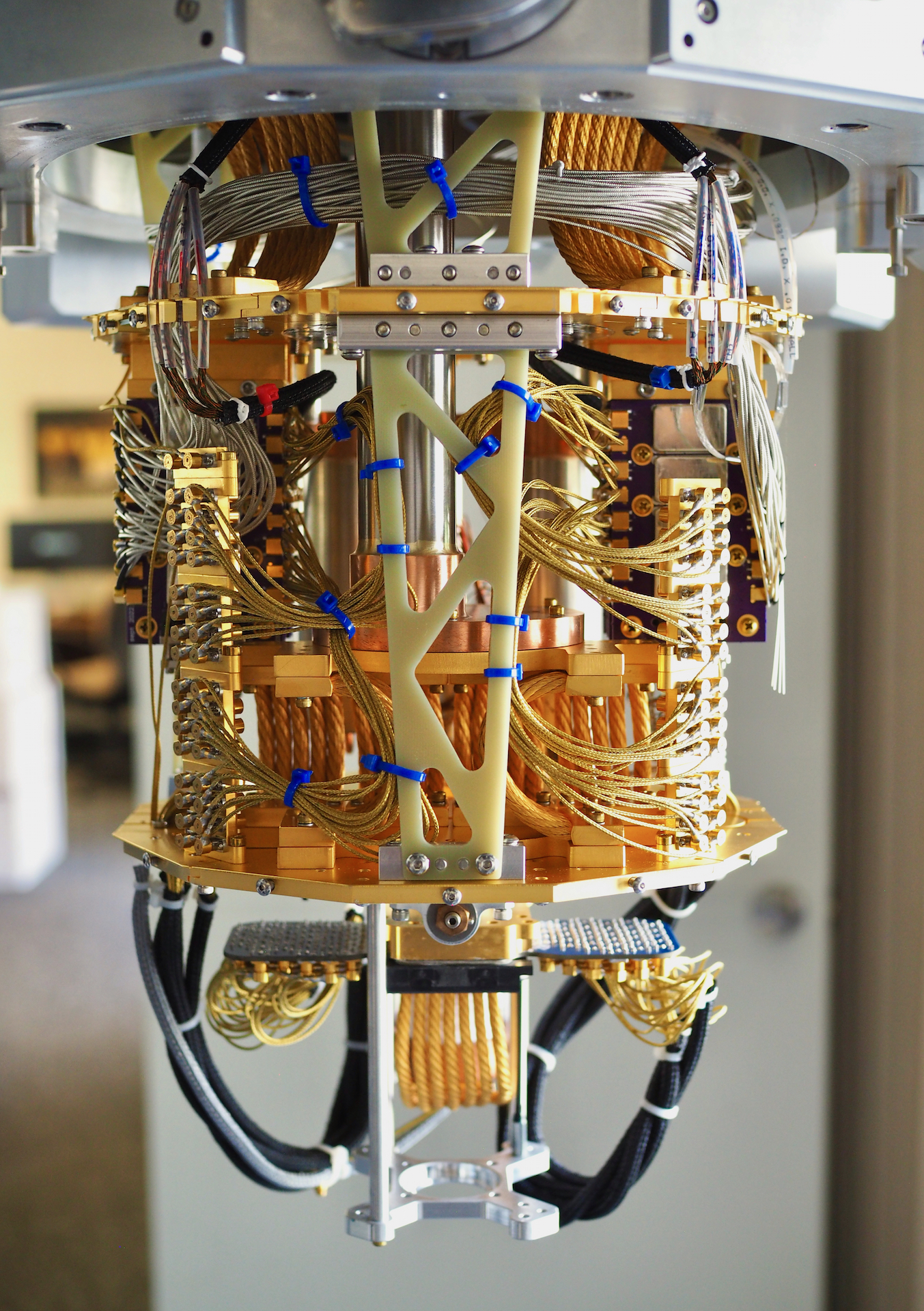}
\caption{\label{fig:rack2}Crysostat internal components including the cryogenic amplifiers.}
\end{figure}

\begin{figure}[t!]
\centering
\includegraphics[scale = 0.25]{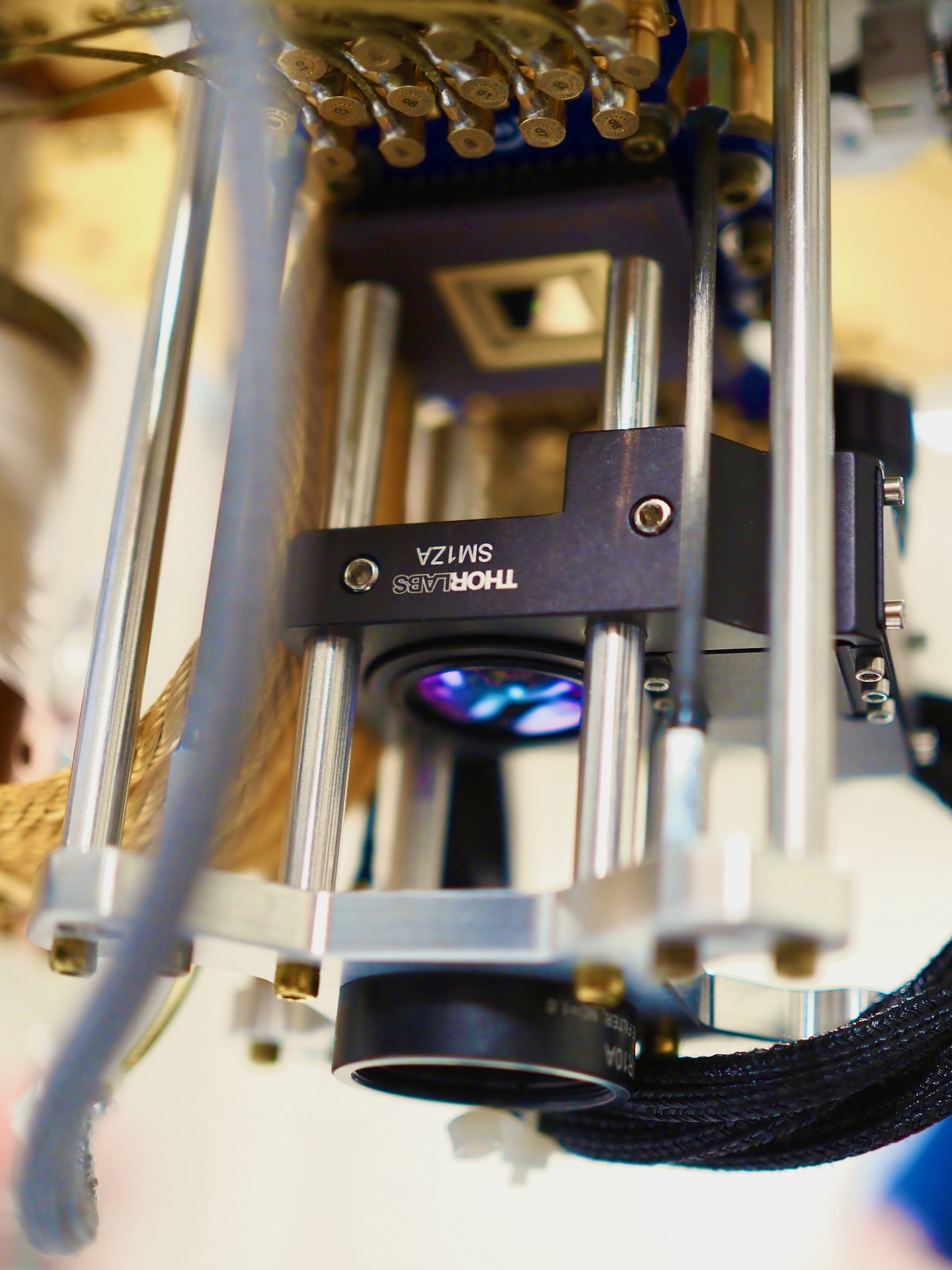}
\caption{\label{fig:rack3}SNSPD array and chip carrier mounted onto the cryostat 1K stage with the final imaging lens mounted on an optical cage system.}
\end{figure}

\begin{figure}[t!]
\centering
\includegraphics[scale = 0.6]{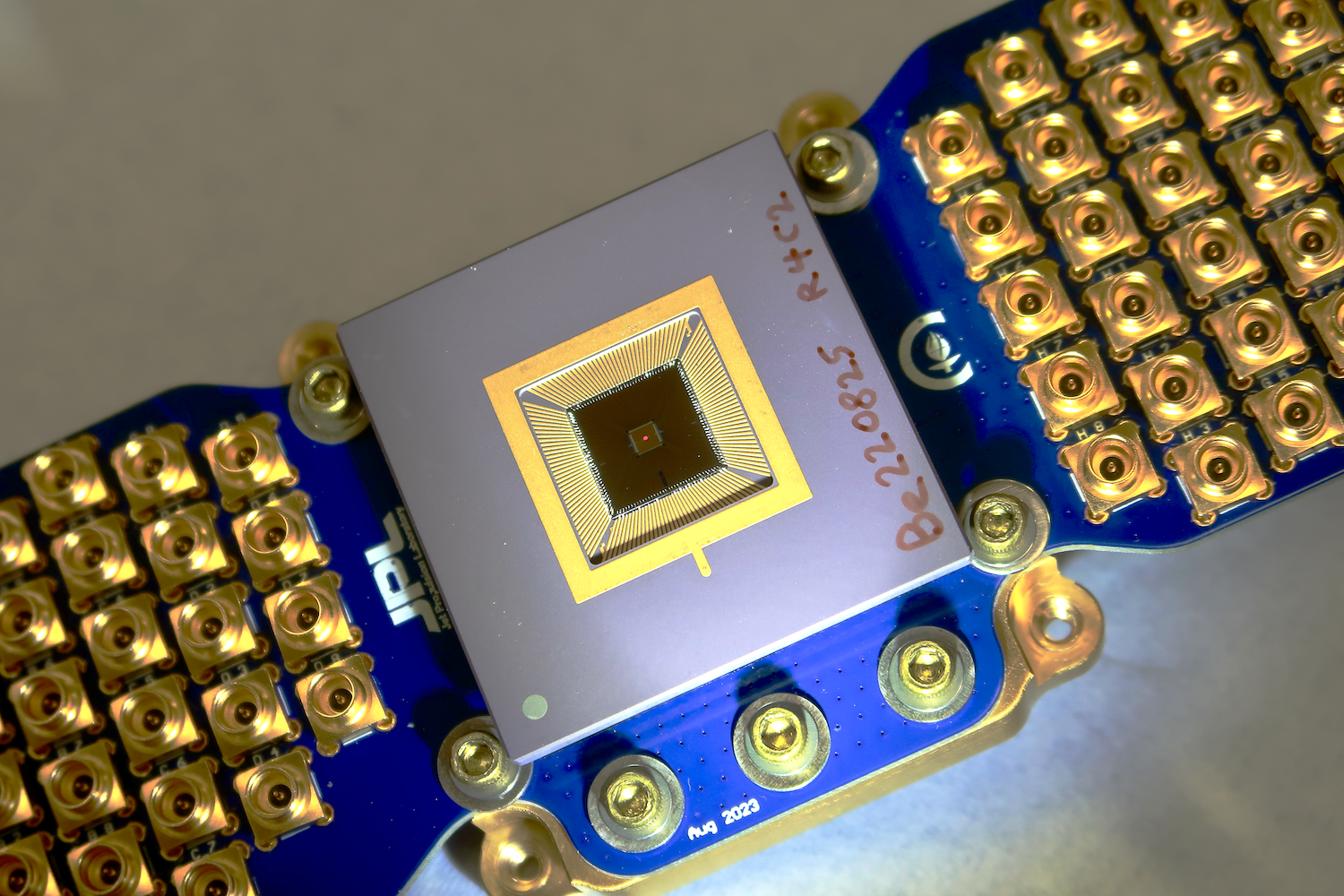}
\caption{\label{fig:rack4}SNSPD array  packaged in a pin grid array (PGA) chip carrier.}
\end{figure}

 Fig.~\ref{fig:rack} shows the array in use with electronic cables leading from the cryostat to the time-to-digital convertor. Fig.~\ref{fig:rack1} shows the rack-mounted lowering mechanism that allows for the cryostat housing to be lowered. Fig.~\ref{fig:rack2} shows the internal components of the crysostat  including the cryogenic amplifiers. Fig.~\ref{fig:rack3} shows the SNSPD array and chip carrier mounted onto the cryostat 1K stage with the final imaging lens mounted on an optical cage system. Fig.~\ref{fig:rack4} shows the SNSPD array packaged in a pin grid array (PGA) chip carrier.

\end{document}